\documentclass[fleqn,usenatbib]{mnras}

\usepackage[T1]{fontenc}

\DeclareRobustCommand{\VAN}[3]{#2}
\let\VANthebibliography\thebibliography
\def\thebibliography{\DeclareRobustCommand{\VAN}[3]{##3}\VANthebibliography}

\usepackage{graphicx}	
\usepackage{amssymb}    
\usepackage{multicol}        
\usepackage{bm}		
\usepackage{pdflscape}	
\usepackage{xcolor} 
\usepackage{ulem}
\usepackage[varvw]{newtxmath}  

\newcommand{\AU}{\rm \,AU}
\newcommand{\degK}{\rm \,K}
\newcommand{\dif}{\mathrm{d}}

\newcommand{\eV}{{\rm \,eV}}
\newcommand{\Fcal}{{\cal F}}

\newcommand{\LJ}{L_{\rm J}}
\newcommand{\Lsun}{L_\odot}
\newcommand{\kB}{\, k_{\rm B}}
\newcommand{\Myr}{\mbox{\,Myr}}
\newcommand{\Mc}{M_{\rm c}}
\newcommand{\Mcore}{M_{\rm c}}
\newcommand{\Mi}{M_{\rm i}}
\newcommand\Msun{M_\odot}
\newcommand{\pc}{\rm \,pc}
\newcommand{\pcc}{{\rm \,cm}^{-3}}

\newcommand{\rc}{r_{\rm c}}
\newcommand{\ri}{r_{\rm i}}
\newcommand{\tff}{t_{\rm ff}}
\newcommand{\vinf}{v_{\rm inf}}

\def\aap{A\& A }

\def\apj{ApJ }
\def\apjl{ApJL }
\def\apjs{ApJS }

\def\mnras{MNRAS }

\def\zap{{ZA} }

\usepackage{xcolor}
\usepackage{ulem}

\definecolor{forestgreen(web)}{rgb}{0.13, 0.55, 0.13}

\title[Multi-scale accretion and delayed massive star formation] {Multi-scale accretion in dense cloud cores and the delayed formation of  massive stars}

\author[E. Vázquez-Semadeni et al.]{
Enrique Vázquez-Semadeni,$^{1}$\thanks{E-mail: e.vazquez@irya.unam.mx}
Gilberto C. Gómez,$^{1}$
and
Alejandro González-Samaniego$^{2}$
\\
$^{1}$Instituto de Radioastronomía y Astrofísica, Universidad Nacional Autónoma de México, 58090, Morelia, Michoacán, México \\
$^{2}$Departamento de Recursos de la Tierra, Universidad Autónoma Metropolitana, Av. de las Garzas No. 10, El Panteón, Lerma de Villada, \\
Estado de México 52005, México
}

\date{Accepted XXX. Received YYY; in original form ZZZ}

\pubyear{2015}

\begin{document}
\label{firstpage}
\pagerange{\pageref{firstpage}--\pageref{lastpage}}
\maketitle

\begin{abstract}
The formation mechanism of massive stars remains one of the main open problems in astrophysics, in particular the relationship between the mass of the most massive stars, and that of the cores in which they form.
Numerical simulations of the formation and evolution of large molecular clouds, within which dense cores and stars form self-consistently, show in general that the cores' masses increase in time, and also that the most massive stars tend to appear later (by a few to several Myr) than lower-mass stars. Here we present an idealized model that incorporates accretion onto the cores as well as onto the stars, in which the core's mass growth is regulated by a ``gravitational choking'' mechanism that does not involve any form of support. This process is of purely gravitational origin, and causes some of the mass accreted onto the core to stagnate there, rather than being transferred to the central stars. Thus, the simultaneous mass growth of the core and of the stellar mass can be computed. In addition, we estimate 
the mass of the most massive allowed star before its photoionizing radiation is capable of overcoming the accretion flow onto the core. This model constitutes a proof-of-concept for the simultaneous growth of the gas reservoir and the stellar mass, the delay in the formation of massive stars observed in cloud-scale numerical simulations, the need for massive, dense cores in order to form massive stars, and the observed correlation between the mass of the most massive star and the mass of the cluster it resides in. Also, our model implies that by the time massive stars begin to form in a core, a number of low-mass stars are expected to have already formed.
\end{abstract}



\section{Introduction}\label{sec1}

The formation mechanism of massive stars and its relationship to the physical nature and state of the dense cores in which these stars form is a crucial ingredient for understanding the origin of the stellar initial mass function (IMF) and the evolution of molecular clouds, which are strongly affected by the feedback from these massive stars. Two main models exist for the formation of massive stars, which are based on fundamentally different scenarios for the process. On the one hand, the {\it competitive accretion} (CA) scenario \citep{Bonnell+01a} assumes that the stars in a forming cluster, as well as the gas from which they accrete, both generate and reside in a common gravitational potential well. Thus, the stars near the bottom of the well accrete at a higher rate, and therefore become more massive than stars in the periphery. On the other hand, the {\it turbulent core} (TC) model \citep{McKee_Tan03} assumes that a massive, dense core must form, so that the pressure within it is high enough to provide the high enough accretion rates onto the protostar that they can persist and resist the feedback from the protostar itself.

In spite of their very different sets of assumptions, they are both developed within the context of a fixed gas mass reservoir. However, recent numerical simulations of star cluster formation under the GHC scenario have shown that the clumps and cores harboring star-forming regions grow in density, mass, and size due to accretion from their respective environments \citep{Heitsch+08, Heitsch_Hartmann08, VS+09, GS_VS20, Camacho+20}, and that the formation of massive stars begins a few to several Myr after the first stars began to form \citep{VS+17}. In addition, \citet{GS_VS20} reported the somewhat surprising result that the dense gas mass and density in the star-forming regions in their simulations manage to continue growing even when star formation has already begun. This implies that somehow the mass transfer rate from the clump-scale to the protostars is {\it not} fully efficient, allowing part of the gas mass to ``stagnate'' in the core, thus causing the latter's mass to grow. In this case, the natural evolution of a core would be to start out as a low-mass star-forming structure, and to evolve into a high-mass one, until finally destroyed by its own stellar population. In addition, this would provide a possible natural, physical explanation for the observed correlation between the mass of the most massive star in a young cluster with the cluster's mass \citep[e.g.,] [] {Weidner_Kroupa06, Weidner+10}, if the latter in turn is somehow capped by the parent clump's mass, as also suggested by \citet{Oey11}.

However, several important questions then arise, such as: 1) What is the mechanism that allows the cores to accumulate mass without transferring it fully to their central parts? 2) How does the core mass growth correlate with the stellar mass growth? 3) At what point during a core's growth does its internal stellar content become capable of disrupting the core? 4) Does this limit depend on the boundary or initial conditions of the core's growth? 

A possible mechanism of core growth could be that turbulence within the core maintains it in approximate equilibrium \citep{McKee_Tan03}, so that it can continue to accumulate mass without collapsing. However, since the core itself must have formed by gravitational contraction and accretion from an external environment, it does not appear feasible that the collapse would be halted by virialization at the core scale while continuing to accrete at the large scale (as suggested, for example, by \citealt{Field+08}), and then resume again to form stars. Instead, a continuous mechanism of core growth would be desirable. An alternative explanation is provided by the {\it Global Hierarchical Collapse} (GHC) scenario \citep{VS+09, VS+19}, which proposes that each level in the hierarchy of density structures within molecular clouds is accreting from its parent structure due to large-scale gravity.

At the scale of accretion onto the young stellar objects (YSOs), it is important to note that the standard Bondi-Hoyle-Lyttleton accretion mechanism \citep{Hoyle_Lyttleton39, Bondi_Hoyle44} assumes that the accretion is driven exclusively by the gravity of the {\it (proto)-stellar} object, neglecting the self-gravity of the gas. Clearly, the gas flow cannot be driven by this mechanism during the {\it prestellar} stage of collapse of a core, since there is no stellar object yet during that period. The gravity from the stellar component also cannot be the driver of the flow at large distances from the stellar object, where the mass is dominated by the gaseous component, even after a YSO has appeared.  Therefore, accounting for the self-gravity of the gas is crucial at all radii during the prestellar stage, and at the large (core) scale, during both stages. This was taken into account in the CA scenario \citep{Bonnell+01a}, although still within the context of a fixed core mass.

In the present paper, we address these questions, along the following steps: We first discuss a plausible mechanism of core mass growth through gravity-driven accretion for isothermal spherical collapse, regulated by the logarithmic slope of the density profile, including the possibility of deviations due to the presence of filaments (Sec.\ \ref{sec:core_growth}). Next, we explore how the instantaneous core mass limits the mass of the most massive star that the core can harbor (Sec.\ \ref{sec:max_stellar_mass}). We then determine the time when the photoionizing  photon flux from the most massive star becomes capable of destroying the core, by equating the power of the accretion onto it to the  heating power from the ionizing photons (Sec.\ \ref{sec:competition}). We conclude in Sec.\ \ref{sec:concls} with a summary of our results and some conclusions. It is worth noting that, in a closely related study, \citet{Myers11} considered both clump-to-core as well as core-to-protostar accretion, although he focused more on the generation of the stellar mass {\it distribution}, while here we focus on the simultaneous evolution of the core's mass and the mass of the most massive possible star.

\section{Core mass evolution} \label{sec:core_growth}

\subsection{Core structure and evolution in strict spherical geometry} \label{sec:core_struc_evol}

The most general analytical solutions of spherical and isothermal gravitational collapse that describe (under idealized conditions) the collapse of dense molecular cloud cores \citep{WS85}, show that the latter go through two distinct asymptotic dynamic evolutionary stages, each of which has two distinct spatial regions, denoted {\it inner} and {\it outer}. This evolution is described via a similarity analysis, in which the variables are nondimensionalized, and are functions of the ``similarity variable'', $x \equiv r/c_{\rm s} t$, where $r$ is the radius, $t$ is time, and $c_{\rm s}$ is the isothermal sound speed.

The time at which the protostar
(formally, the {\it singularity}) forms is usually denoted $t=0$, and so the time interval $t<0$ (which implies $x<0$ as well) corresponds to the {\it prestellar} stage, while the interval $t>0$ corresponds to the {\it protostellar} stage. During the prestellar stage, the inner region has a uniform density and a (negative) infall speed that scales linearly with radius ($v(x) \propto x$), so that its magnitude decreases inwards. The size of the inner region is comparable to the Jeans length corresponding to its uniform density \citep{Keto_Caselli10}. The outer region has a density profile $\rho(x) \propto \lvert x\rvert^{-2}$ and an infall speed that decreases with radius as $v(x) \propto x^{-1}$ if the speed is required to vanish at infinity (see eq.\ [3.8] of \citealt{WS85} with $y_\infty = 0$). During the {\it protostellar} stage, the inner region has $\rho(x) \propto x^{-3/2}$ and $v(x) \propto -x^{-1/2}$, and the outer region has $\rho(x) \propto x^{-2}$ and again $v(x) \propto -x^{-1}$. 

In the strict spherical description, during the prestellar stage, the mass contained within a certain fixed radius increases because both the central and the mean density increase with time. However, during the protostellar stage, the {\it gaseous} mass in the core remains fixed, because it can be shown that the radial density profile $\rho(x) \propto x^{-2}$ and the nondimensionalization of $\rho$ by the time-dependent quantity $4 \pi G t^2$ combine to make the density profile independent of $t$. That is, the density profile becomes fixed in time, and so does the gaseous mass in the core \citep[see also][]{Murray_Chang15}. Only the mass of the central protostar increases with time. 

\begin{figure}
\centering
\includegraphics[width=0.5\textwidth]{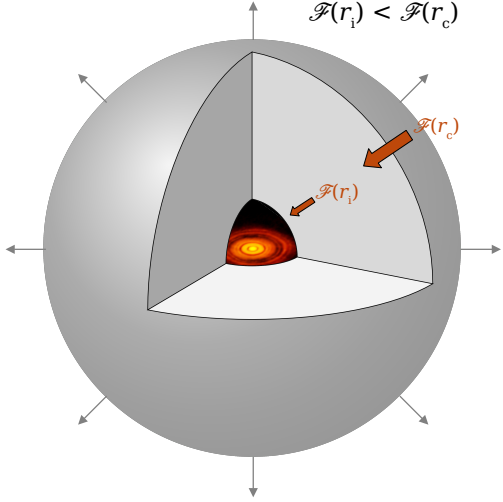} 
\caption{
Schematic diagram of the assumed core boundaries and structure. The core is assumed to be immersed in a medium with a uniform (or slowly-varying) density, into which the core merges at an outer radius $\rc$, whose initial value is $r_0$. In its innermost region, the core is assumed to end at an internal radius $\ri$, below which the spherical symmetry ends, perhaps due to the presence of an accretion disk. We refer to the region $r < \ri$ as the {\it stellar region}. The core accretes mass from its environment at a rate $\Fcal(\rc)$, and accretes {\it onto} the stellar region at a (generally lower) rate $\Fcal(\ri)$. We refer to the condition $\Fcal(\ri) < \Fcal(\rc)$ as {\it gravitational choking}.
} 
\label{fig:choking}
\end{figure}

In the remainder of this section, we revisit the accretion flow into and across the core towards the star,  using the prescription introduced in \citet[] [hereafter Paper I] {Gomez+21}, where we presented an approximate model for the evolution of the {\it average} logarithmic slope of the radial density profile of a core embedded in a uniform background.\footnote{Since the background is assumed uniform, or at least shallower than the core, it must also be gravitationally unstable, and therefore, undergoing collapse. However, as proposed in the Global Hierarchical Collapse scenario \citep{VS+19}, the background may be undergoing large-scale collapse towards a {\it different} collapse center than that of our core, so that the latter can be thought of as collapsing locally while ``riding'' the large-scale collapse flow in a {\it conveyor belt} fashion \citep{Longmore+14}.} The model can be thought of as describing the temporal transients undergone by the density profile during the early stages of the collapse, as well as the  ``spatial transients'' (flattenings) of the slope happening at the outer and inner boundaries of the core. The outer edge, denoted $\rc$, is defined as the radius at which the core merges with the background, while the inner edge, denoted $\ri$, is defined as the radius where the spherical symmetry assumption must break down due to the formation of an accretion disk (Fig. \ref{fig:choking}). Note that these transients are not considered in the classical asymptotic Larson-Penston \citep{Larson69, Penston69} solutions, which, by their very asymptotic nature, are valid only far away from temporal transients and border effects. In reality \citep[for example, in numerical simulations including a background; e.g.,] [] {Naranjo+15}, the slope varies with time and radius, and the model approximates it by a single, time-dependent slope from $\ri$ to $\rc$. The model allows the calculation of the gravity-driven accretion rate at any given radius $r$, which we will compute at both $\rc$ and $\ri$ to respectively estimate the accretion {\it onto the core} and {\it from the core onto its innermost parts}, where it is assumed to form stars.

It is worth noting that the evolving power-law profile of our model represents an ``average'' slope of the formal solution by \citet{WS85} only in a loose, qualitative way, since our model is not ``derived'' by any form of averaging from the formal solution. Instead, our model simply represents an approximate, idealized spherical core, forced to have a power-law profile from its center to a radius equal to the initial Jeans length in the background medium, $r_0$, and accreting  at each radius at the rate dictated by the gravity internal to that radius. In this sense, at the start of its evolution (formally $t\sim -\tff$, where $\tff$ is the free-fall time of the background medium, so that $t=0$ when the singularity forms), our model represents exclusively the flat part of the true density profile. Subsequently, during the time interval $-\tff \lesssim t < 0$, the flat part in the true solution shrinks in size, and the $r^{-2}$ region grows in size. Our $r^{-p}$ profile, with $0 < p < 2$, qualitatively represents the transition from a fully flat to a fully $r^{-2}$ profile, albeit with no detailed averaging of the true profile.

Finally, we remark that the fundamental aspect of the present model is that the core's mass is time-dependent, due to accretion onto it from its environment. Therefore, the core's mass grows simultaneously with the stellar mass it contains, and the main goal of this paper is to compare these two growth rates.


\subsection{Core mass evolution in spherical geometry. Gravitational choking} \label{sec:sph_core_growth}

Already the first numerical simulations and analytical calculations of isothermal spherical collapse showed the development of a density profile approaching $\rho (r) \propto r^{-2}$ \citep{Larson69, Penston69}. The early interpretation of such a profile was that a period of quasi-static contraction \citep{Shu77} was required in order for sound waves to establish detailed pressure balance throughout the core, similarly to the case of a hydrostatic Bonnor-Ebert \citep{Ebert55, Bonnor56} sphere. However, a recent study \citep{Li18} has shown that the $r^{-2}$ profile can arise simply from letting the radial infall speed at every radius $r$ be the gravitational velocity induced by the gas mass internal to that radius,
\begin{equation}
\vinf(r) = \sqrt{\frac{2GM(r)}{r}},
\label{eq:vinf_eq_vgrav}
\end{equation}
where
\begin{equation}
M(r) = \int_0^r 4 \uppi \rho(r') r'^2 \dif r',
\label{eq:M_of_r}
\end{equation}
and requiring that the  accretion rate at radius $r$
\begin{equation}
\Fcal(r) = 4 \uppi \rho(r) v(r)r^2
\label{eq:accr_rate}
\end{equation}
be independent of radius.  Note that $\Fcal(r)$ has units of mass per unit time, and corresponds to the rate at which mass accretes through a surface of radius $r$.

Furthermore, in Paper I an additional step was taken by considering the transient evolution of the logarithmic slope of the core's density profile. The core was assumed to begin its life as a moderate, arbitrary density fluctuation, in an isothermal medium, of radius $\rc \approx \LJ(\rho_0)/2$, where  $\rho_0$ is the density of the background medium and $\LJ(\rho_0)$ is the Jeans length at that density.
During the prestellar stage studied in that paper, the core was assumed to evolve by increasing the slope of its density profile, keeping $\rc$ fixed. However, as we shall see below, $\rc$ can vary over time, and so in general we will have $\rc = \rc(t)$. For the radius-averaged density profile from $r=0$ to $r=\rc$, Paper I assumed a power law of the form 
\begin{equation}
\rho(r) = \rho_0 \left(\frac{r}{\rc}\right)^{-p}
\label{eq:dens_prof}
\end{equation}
at all times. Note that this is not strictly true since, as seen in eq.\ \eqref{eq:dpdt} below, the slope varies at different rates at different radii. In this context, $p$ should be regarded as the {\it mean} logarithmic slope of the density profile over the core's radial extent. 

Then, making the approximations that the infall speed is given by eq.\ \eqref{eq:vinf_eq_vgrav} and that the density profile is given by eq.\ \eqref{eq:dens_prof} at all times (i.e., that it evolves from one power law to another), and introducing them into the continuity equation, Paper I showed that the rate of change of the logarithmic slope $p$ at radius $r$ is given by
\begin{equation}
    \frac{\dif p(r)}{\dif t} = \left( 3-\frac{3}{2}p\right)\left( \frac{4\uppi G \rho_0}{3-p}\right)^{1/2} \left[ \frac{(r/\rc)^{-p/2}}{-\ln(r/\rc)}\right].
  \label{eq:dpdt}
\end{equation}

Furthermore, Paper I noted that the sign of this derivative is determined exclusively by the factor $(3 -3p/2)$ in the right-hand side of eq.\ \eqref{eq:dpdt}, so that, if $p<2$, then the slope increases over time, while if $p>2$, then the slope decreases. If $p=2$, the slope remains stationary. That is, $p=2$ is an {\it attractor} for the logarithmic slope of the density profile of a flow generated by gravitational attraction of the internal mass, {\it under spherical symmetry}.

An additional implication of equations \eqref{eq:vinf_eq_vgrav} and \eqref{eq:dens_prof} for the velocity and density profiles is that, if $p \ne 2$, then the  accretion rate given by \eqref{eq:accr_rate} is {\it not} constant with radius, but instead depends on $r$ as 
\begin{equation}
\Fcal(r) = \left(\frac{128 \uppi^3 G \rho_0^3 \rc^{3p}} {3-p} \right)^{1/2} r^{\left(3 - \frac{3}{2}p\right)},
\label{eq:acc_rate_of_r}
\end{equation}
As a reference, the above expression, expressed in dimensional form and evaluated at the initial outer boundary of the core, $r_0 = \rc(t=0)$, reads
\begin{equation}
    \Fcal(r_0) = \frac{1.88 \times 10^3 (\Msun \Myr^{-1})}{(3-p)^{1/2}}
                  \left( \frac{r_0}{1 \pc} \right)^3
                  \left( \frac{n_0}{10^3\pcc} \right)^{3/2}
  \label{eq:mass_flux_dim}
\end{equation}

Differentiating eq.\ \eqref{eq:acc_rate_of_r} with respect to the radius then gives
\begin{equation}
\frac{\dif \Fcal(r)} {\dif r} = \frac{3(2-p)} {2} \left(\frac{128 \uppi^3 G \rho_0^3 \rc^{3p}} {3-p} \right)^{1/2} r^{\left(2 -\frac{3}{2}p\right)}.
\label{eq:acc_rate_rad_grad}
\end{equation}
This equation shows that {\it the  accretion rate $\Fcal(r)$ decreases with decreasing radius for} $p<2$, implying that not all of the mass entering the core at its outer boundary can be transferred to its center. Some of the mass is trapped in the core, causing the core's mass to grow, as long as $p < 2$. We refer to this phenomenon as {\it gravitational choking} of the gravity-driven inflow. It is important to note that this phenomenon does not involve any kind of support;  the gas is still freely infalling at all radii as
dictated by eq.\ \eqref{eq:accr_rate}, but the gravity-driven infall rate decreases with decreasing radius. That
is, for a profile with $p<2$, the gravity of the material inside $r$ just cannot transfer mass at a constant rate across the radial extent of the core.

Finally, it is also important to note that eq.\ \eqref{eq:acc_rate_of_r} implies, for strict spherical symmetry, that the mass accretion flow vanishes at $r=0$ during the {\it prestellar} stage, in which $p < 2$. 
This situation changes during the {\it protostellar} stage, which starts when the density profile slope reaches $p=2$, since at this point, {\it the  accretion rate $\Fcal(r)$ becomes independent of radius}. That is, during the protostellar stage, {\it in spherical geometry}, all of the mass entering through the core's boundary is transferred to the central stellar object.

Summarizing, under spherical symmetry, three important conditions are reached when the slope reaches the value $p=2$:

\begin{itemize}

\item The slope becomes stationary.

\item The mass accretion rate becomes independent of radius (all the mass entering the core on the outside is uniformly transferred across all radii).

\item A singularity (protostar) is formed.

\end{itemize}

Note that the latter condition does not follow from the model in Paper I, but rather from the similarity solutions and numerical simulations \citep[e.g.][]{Larson69, WS85}, which show that the entire density profile becomes a single power law with a logarithmic slope of $-2$ when the central singularity first appears. The exact solution has a constant-density inner region and an $r^{-2}$ outer envelope during the prestellar stage. When the central region shrinks to zero radius, the singularity appears.


\subsection{Mass fraction retained in the core per unit time} \label{sec:retained_mass_frac}

We now wish to estimate the amount of mass that is retained in the core and the amount of mass that goes into ``stars'' as a function of time. However, as noted above, eq.\ \eqref{eq:acc_rate_of_r} implies that, for all $p < 2$, the  accretion rate $\Fcal(r)$ vanishes at the center. We can circumvent this problem by  noting that the spherical geometry assumption cannot strictly extend to zero radius, but rather must end at some inner radius $r_{\rm i}$, representing, for example, the radius at which an accretion disk forms.\footnote{Note that this internal boundary of the core is not related in any way to the transition between the inner and outer regions of the core discussed in Sec.\ \ref{sec:core_struc_evol}. Instead, it just accounts for the fact that the spherical symmetry must break down at radii comparable to the size of an accretion disk.} In this case, the fraction of mass retained in the core per unit time is given by the accretion rate at the outer boundary, $\rc$, minus the accretion rate at $r_{\rm i}$. We can thus write
\begin{equation}
\dot M_{\rm core} = {\cal F}(\rc) \left[1 - \frac{{\cal F}(r_{\rm i})} {{\cal F}(\rc)} \right] = {\cal F}(\rc) \left[1 - \left(\frac{r_{\rm i}} {\rc} \right)^{\frac{3}{2} (2-p)} \right].
\label{eq:dot_Mcore_of_r}
\end{equation}
%
%
%

It is worth noting that, in the idealized, perfectly spherical collapse, during the prestellar stage with $p < 2$, {\it all} of the mass entering the core is retained in the gaseous phase, since the central density has not diverged yet, and so there is zero mass transfer to the ``stars'' (the region $r < \ri$). Conversely, the mass transfer from the boundary to the center becomes 100\% efficient at the onset of the protostellar stage (the time at which the singularity---the protostar---appears), because at that time the logarithmic slope becomes $p = 2$.  That is, in the idealized, perfectly spherical case, the core transitions from having zero net mass transfer to its center to having 100\%-efficient mass transfer at the moment of singularity formation.

\begin{figure}
\centering
\includegraphics[width=0.49\textwidth]{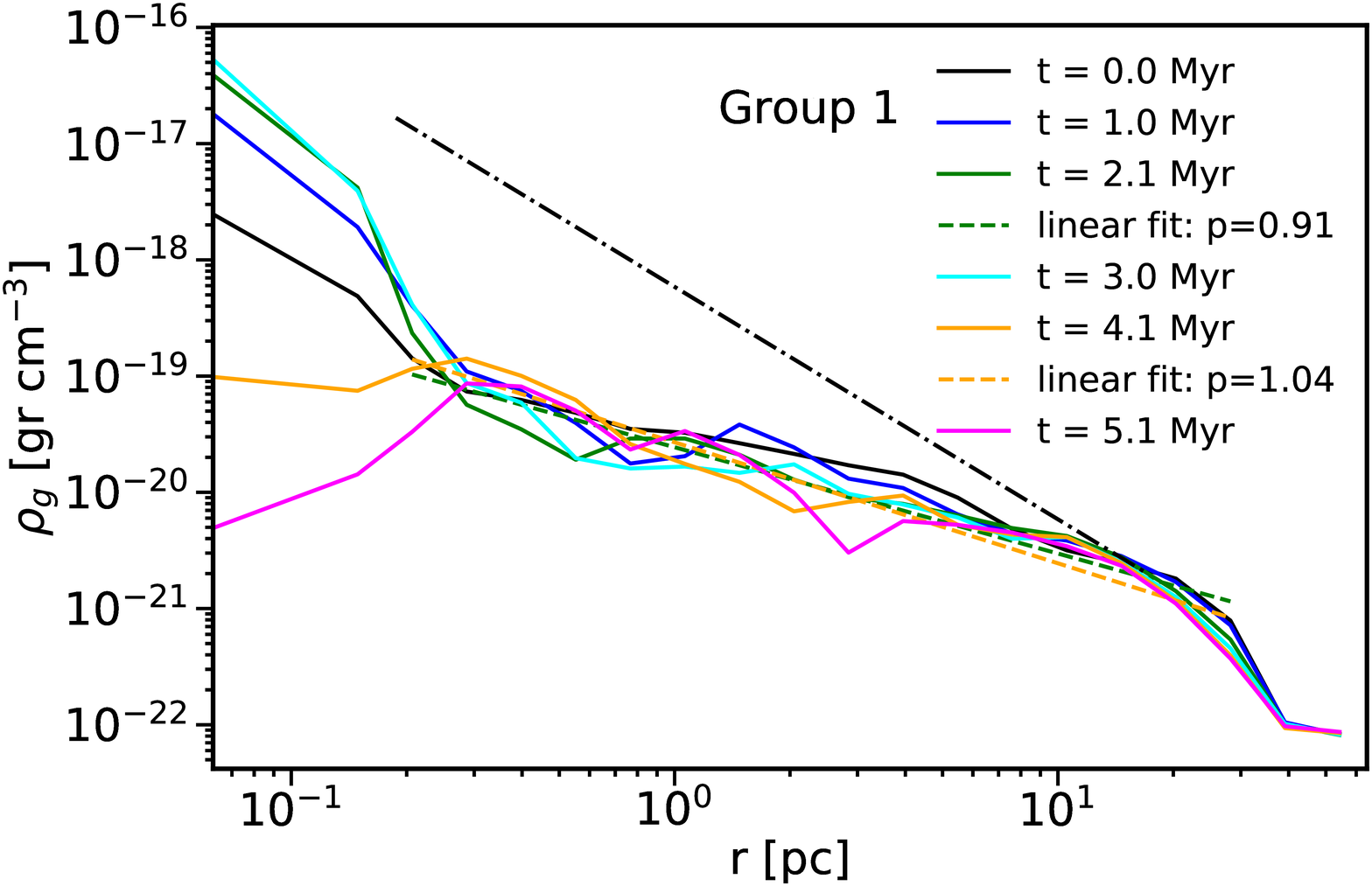} 
\includegraphics[width=0.49\textwidth]{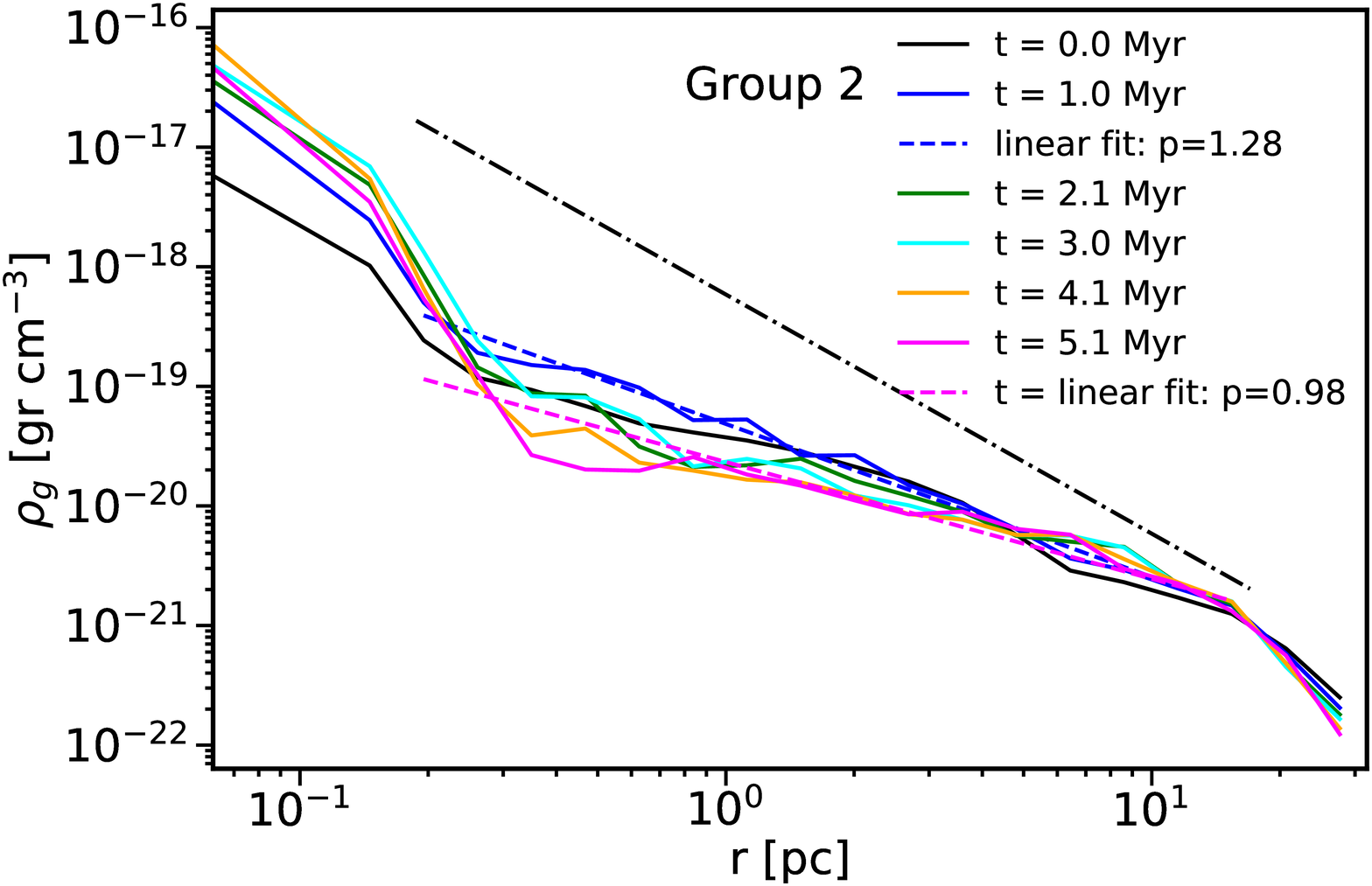}
\includegraphics[width=0.49\textwidth]{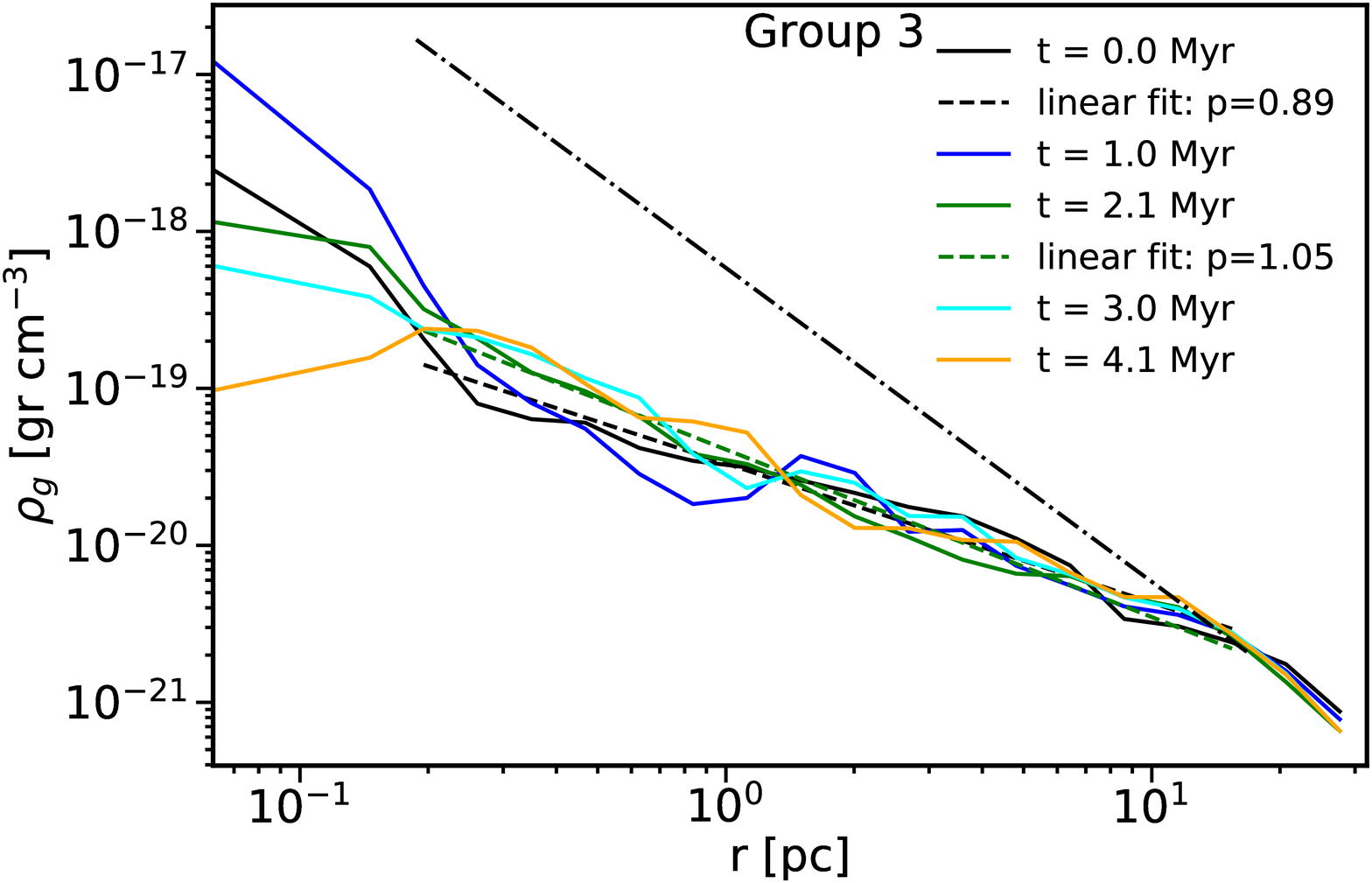}
\caption{Spherically-averaged density profile for ``clumps'' in the numerical simulation LAF0 of \citet{GS_VS20} at different times after the formation of the first stellar particle; each clump is clearly identified as a different star-forming region in the numerical box and they are labeled as Group 1, Group 2, and Group 3. The median of density cells in spherical bins (shells) is used to compute the density profile. In each panel dashed lines represent the shallowest and steepest slopes for the different times considered. In each clump and at all the times considered, density profile slopes are always lower than $p=2$ (represented by the dot-dashed line in each panel). Interestingly, we do not see an evolutionary trend in the profiles.
}
\label{fig:profile_ART}
\end{figure}

However, in real molecular cloud cores, the actual profiles appear to be shallower than $-2$ {\it even after protostellar objects have appeared}. Indeed, the mean slope for the compilation of cores examined in Paper I was found to be $p \approx 1.9$ for low-mass cores and $p \approx 1.7$ for high-mass cores. Moreover, in Fig.\ 
\ref{fig:profile_ART} we show the spherically-averaged density profiles of the three star-forming regions appearing in the simulation LAF0 (without stellar feedback) studied by \citet[][see Appendix \ref{sec:methods}]{GS_VS20}, from the onset of star formation to a few megayears later, showing that the logarithmic slopes are consistently shallower than $-2$.\footnote{It is important to note that these slopes cannot be attributed to the profiles being determined by the gravity of the central stellar objects, as in Shu's inside-out collapse solution \citep{Shu77}, since the gaseous mass at most of the radii indicated in the figure is much larger than the stellar mass \citep[see, e.g. Fig.\ 3 of] [] {GS_VS20}.} The reason for this is not clear, but one possibility is that filamentary accretion flows may cause a flattening of the spherically-averaged density distribution. As a proof-of-concept, in Appendix \ref{sec:fils_dens_profile} we show that the superposition of a uniform-density filament on top of a power-law spherical density distribution flattens the net spherically-averaged profile, thus decreasing the depth of the core's gravitational potential well.

We thus suggest that the presence of filaments in hub-filament systems can flatten the spherically-averaged density profile, thus causing the gravitational attraction of the inner gas mass to be insufficient for driving all of the material entering the core to be transferred to the central stars, and therefore allowing the core to grow by mass accumulation. In the next section we now describe the simultaneous mass growth of the central star(s) and of the core, as a function of time and of the density profile's slope.

%
%

\subsection{Evolution of the core's mass and radius} \label{sec:growth}

\begin{figure}
\centering
\includegraphics[width=0.5\textwidth]{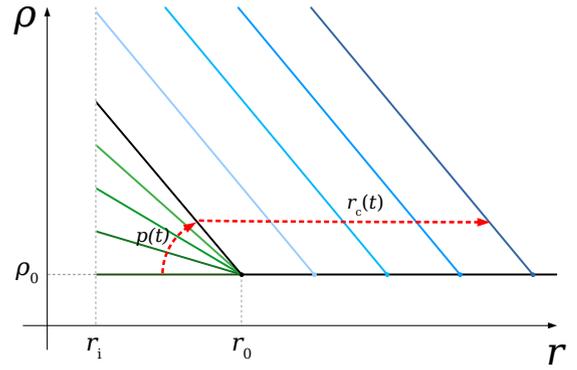} 
\caption{Evolution of the core's density profile. Since the core results from the collapse of a region at constant density, its outer edge ($\rc$) is initially equal to the environment's Jeans length. As the gas surrounding the core is accreted, the core's density profile steepens. After the density slope becomes stationary, the continued accretion implies that the core's size increases (see appendix \ref{sec:app_acc_times}). The core's inner boundary ($\ri$) is assumed to remain constant throughout this evolution.
}
\label{fig:evolution}
\end{figure}

Following Paper I, let us assume a core with the density profile given by equation \eqref{eq:dens_prof}, surrounded by an environment at constant density $\rho_0$. Let us ignore, for the moment, the core's inner boundary at $\ri$ and label as ``core'' the gas with density greater than $\rho_0$. Then, the core's mass $\Mcore$ is given by
%
%
\begin{align} \label{eq:Mcore}
    \Mcore(p) &= \frac{4\uppi}{3-p} \rc^3 \rho_0 \\
              &= \frac{732 \Msun}{3-p} 
                  \left( \frac{\rc}{1\pc} \right)^3 \left( \frac{n_0}{10^3 \pcc} \right). \nonumber
\end{align}
This equation gives the time dependence of the core's mass implicitly, through the time dependence of $p$ given by eq.\ \eqref{eq:dpdt}.
However, since the slope $p$ becomes stationary after forming a star (possibly at a value $p < 2$ due to the presence of filaments), and we are assuming that $\rho_0$ remains constant during this stage, the mass growth of the core caused by accretion and gravitational choking requires the core's radius $\rc$ to increase---i.e., the core expands (see Fig. \ref{fig:evolution}). This implies a departure from the asymptotic self-similar solution, which implicitly pushes the core boundaries to infinity. For a finite core, then, the expansion rate $\dif \rc/\dif t$, after the slope becomes stationary, can be related to the mass growth rate within the core by taking the time derivative of eq. \eqref{eq:Mcore},
\begin{equation}
    \frac{\dif \Mcore}{\dif t} = \frac{12 \uppi \rho_0 \rc^2}{3-p} \frac{\dif \rc}{\dif t}.
    \label{eq:Mdot_core}
\end{equation}
%


In turn, this mass growth rate must be equal to the net  accretion rate given by eq.\ \eqref{eq:dot_Mcore_of_r}, neglecting the factor within square brackets, since we are ignoring the core's inner boundary.
We thus have:
\begin{equation}
\frac{\dif \rc} {\dif t} \approx \frac{2} {3} \left[ 2 \left(3-p\right)\uppi G \rho_0\right]^{1/2} \rc 
\label{eq:drcdt}
\end{equation}
It is thus seen that 
$\rc$ grows approximately exponentially.

In appendix \ref{sec:app_acc_times} we show that the mass increase due to the evolution of $p$ occurs on a shorter timescale than that for the increase of $\rc$. Therefore, the core's mass must grow initially by increasing the slope of its density profile at nearly constant radius $\rc$ (during the prestellar stage), and later by increasing its size at a nearly fixed slope (during the protostellar stage). However, due to the presence of filaments, we expect that the slope may saturate at a value $p < 2$. Thus, the core's mass can continue to grow by gravitational choking until the accretion supply is exhausted, and the evolution of the core's mass can be followed together with that of the internal stellar object(s).

We now take into account the fact that the stellar mass accumulates at the center. In this case, eq.\ \eqref{eq:drcdt} reads
\begin{equation}
\frac{\dif \rc} {\dif t} \approx \frac{2} {3} \left[ 2 \left(3-p\right)\uppi G \rho_0\right]^{1/2} \rc \left[1 - \left(\frac{\ri}{\rc}\right)^{3(2-p)/2} \right],
\label{eq:Mdot_eq_mass_flux}
\end{equation}
and the gravitational velocity $\vinf$ at radius $r$ is given by 
\begin{equation}
\vinf(r) \approx \sqrt{\frac{2 G M_{\rm tot}}{r}} = \sqrt{\frac{2 G \left[M_{\rm g}(r) + \Mi\right]}{r}},
\label{eq:vgrav_tot}
\end{equation}
%
where $M_{\rm g}(r)$ is the gas mass contained between $\ri$ and $r$ and $\Mi$ is the total mass that has been accreted onto the internal ``stellar'' region through $\ri$ over the entire evolution. We refer to $\Mi$ as the ``stellar mass''.

At short radii, where $r \rightarrow \ri$, we have that $\Mi \gg M_{\rm g}$, and so
\begin{equation}
\vinf(\ri) \approx \sqrt{2 G \Mi}\, \ri^{-1/2}.
\label{eq:vgrav_ri}
\end{equation}
%
Therefore the mass accretion rate (the surface-integrated mass flux) into the ``stellar region'', $\Fcal(\ri) = 4 \pi \rho(\ri) \vinf(\ri) \ri^2$, using the density profile given by eq.\ \eqref{eq:dens_prof}, becomes
\begin{equation}
\Fcal(\ri) = \left(32 \uppi^2 G \Mi\right)^{1/2} \rho_0 \rc^p \ri^{\frac{3}{2} -p}
\label{eq:dotm_ri}
\end{equation}%

On the other hand, at the outer radius of the core, $\rc \gg \ri$, we have $\Mi \ll M_{\rm g}$, and therefore the infall speed is
\begin{equation}
\vinf(\rc \gg \ri) \approx \left(\frac{8 \uppi G \rho_0}{3-p}\right)^{1/2} {\rc},
\label{eq:vgrav_rc}
\end{equation}
%
and the accretion rate onto the core is then
\begin{equation}
\Fcal(\rc) = \left( \frac{128 \uppi^3 G}{3-p} \right)^{1/2} \rho_0^{3/2} \rc^3.
\label{eq:dotm_rc}
\end{equation}

It is important now to determine under which conditions can the accretion from the core to the stellar region overcome the accretion onto the core. For this, using eqs.\ \eqref{eq:dotm_ri} and \eqref{eq:dotm_rc}, we compute the ratio of the inner and outer accretion rates:
\begin{equation}
\frac{\Fcal(\ri)}{\Fcal(\rc)} \propto 
    \left(\frac{\Mi} {\rho_0} \right)^{1/2}\, \ri^{\frac{3}{2}-p} \rc^{p-3}.
\label{eq:dotm_ratio}
\end{equation}
This expression can be more easily interpreted noting that $\Mi/\rho_0 = (\Mc/\rho_0) (\Mi/\Mc)$ and that, from eqs.\ \eqref{eq:M_of_r} and \eqref{eq:dens_prof}, 
\begin{equation}
\frac{\Mc}{\rho_0} = \frac{4 \pi \rc^3}{3-p}.
\label{eq:M_over_rhoo}
\end{equation}
%
Therefore, writing $\ri = \epsilon \rc$, we finally obtain
%
\begin{equation}
\frac{\Fcal(\ri)}{\Fcal(\rc)} \propto \epsilon^{\frac{3}{2}-p}.
\label{eq:dotm_ratio_norm}
\end{equation}
%
This equation shows that, for $p > 3/2$, the accretion rate ratio {\it increases} as the size of the inner stellar region becomes a smaller fraction of the core's size. This happens precisely as the core's radius increases, therefore allowing for the possibility that the core is {\it depleted} by the accretion onto the stellar region if $p > 3/2$. This depletion implies that it is possible for the mass accumulated in the stellar region to become larger than the gaseous mass of the core. 
On the contrary, for $p < 3/2$ the core's mass always grows faster than the mass of the stellar region. 

\begin{figure*}
\centering
\includegraphics[width=1.\textwidth]{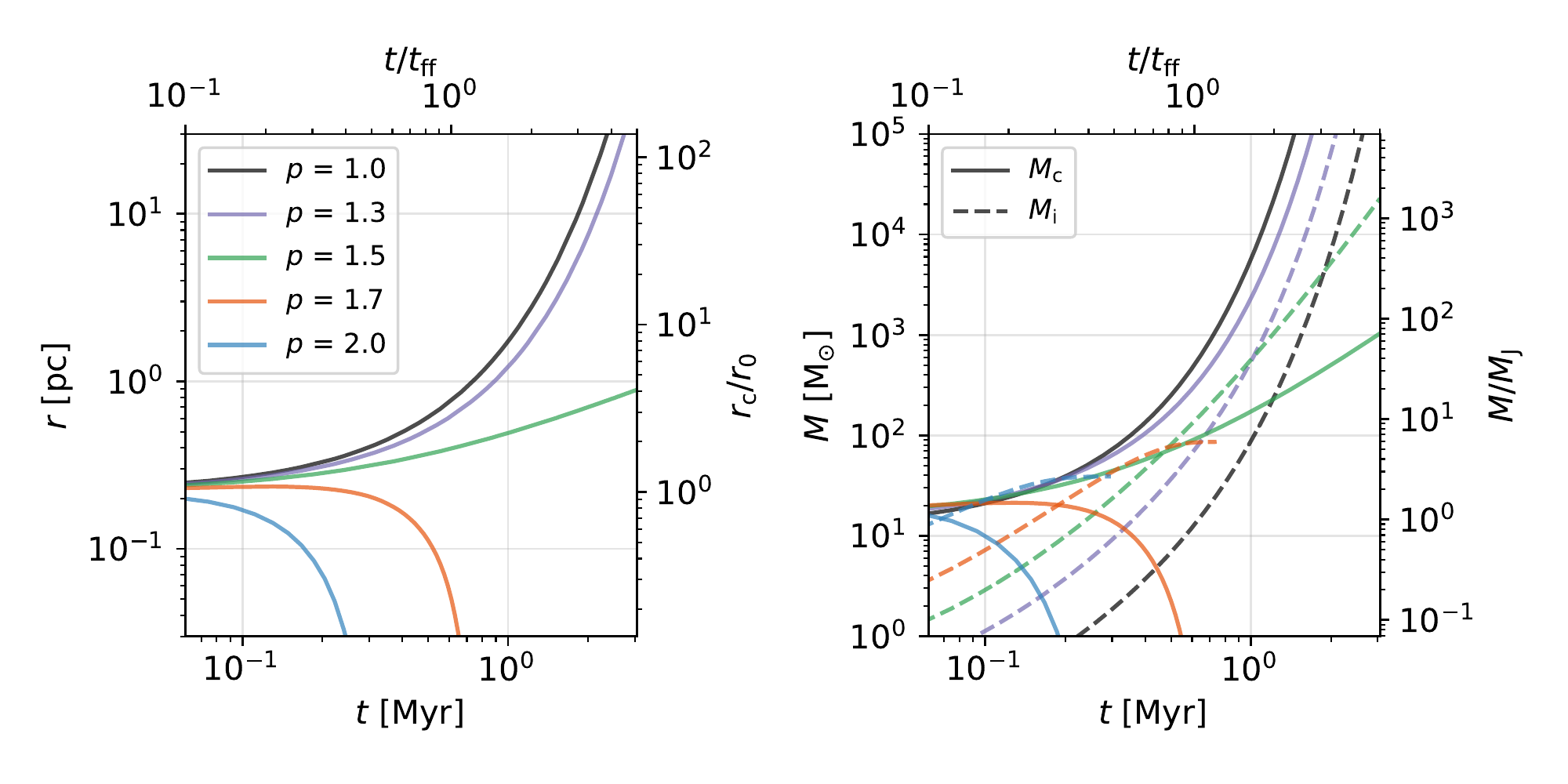}
\caption{Core size ($\rc$, {\it left}), and gaseous and stellar mass ($\Mc$ and $\Mi$, {\it right}) as a function of time for a range of $p$-values. The masses are obtained by numerically integrating the fluxes at $\rc(t)$ and the core's inner boundary (at $\ri$). When $p > 3/2$, the flow across $\ri$ is larger than the flow across $\rc$. So, the core ends up being depleted. In these plots, we have assumed an environmental density $\rho_0 = 3\times 10^3\pcc$ and an initial radius of the core $\rc(t=0)$ equal to the Jeans length at $\rho_0$.} 
\label{fig:Mcore}
\end{figure*}

Therefore, $p = 3/2$ is another critical value of the logarithmic slope, determining whether the core grows or is depleted by the accretion onto the stellar region. This is illustrated in Fig.\ \ref{fig:Mcore}, which shows the evolution of the core size ({\it left} panel), and of the core's gas mass and the stellar mass ({\it right} panel) obtained by numerically integrating the mass fluxes\footnote{By numerically integrating the mass fluxes, and so calculating the core and stellar mass evolution, we avoid the need to make assumptions about the masses relative importance in the estimation of infall velocities, as it was necessary leading to eqs. \eqref{eq:vgrav_ri} and \eqref{eq:vgrav_rc}.} ${\cal F}(\ri)$ and ${\cal F}(\rc)$ in time for a range of logarithmic slope values. The surrounding medium is assumed to have a density of $3 \times 10^3 \pcc$ and temperature of $15 \degK$, implying a Jeans radius of $0.22\pc$, which is taken as the initial value for the core's radius. The core's inner boundary is assumed to be at $\ri = 3\times 10^3 \AU$. We choose this value of $r_{\rm i}$ as representative of the region where an accretion disk begins to form, based on estimates of the Oort cloud's mean radius \citep{Morbidelli05}.  As expected, for $p > 3/2$ the flow across the inner boundary depletes the gaseous mass of the core, although the stellar mass still becomes larger than the core's mass for $p=3/2$.


\section{The most massive star a core can harbor} \label{sec:max_stellar_mass}

\subsection{The competition between core and stellar mass growth} \label{sec:competition}

Having obtained the mass accretion rate onto the core, we can now estimate the mass of the most massive star the core can harbor without its accretion supply being destroyed by the UV feedback from this star.\footnote{We consider the UV flux rather than the radiation pressure because we are interested in the competition between the stellar feedback and the accretion onto {\it the core}, which occurs at scales larger than the core scale ($\gtrsim 0.1$ pc). At those scales, the dominant form of continuous feedback is already UV radiation \citep[e.g.,] [] {Sales+14}.} Assuming that the gas infall velocity onto the core at every radius is given by the free-fall velocity (eq.\ [\ref{eq:vinf_eq_vgrav}]) driven by the total (stellar and gaseous) mass internal to that radius, then the power $P_K$ associated to the kinetic energy density in the accretion flow at the boundary, $K = 1/2\ \rho(\rc) \vinf^2(\rc)$,  is
\begin{align}
    P_K &= 4 \uppi \rc^2 K \vinf(\rc) \\
        &\sim \frac{0.97 \Lsun}{(3-p)^{3/2}}
           \left( \frac{\rc}{1 \pc} \right)^5
           \left( \frac{n_0}{10^3 \pcc} \right)^{5/2}, \nonumber
\end{align}
where the order-of-magnitude value in the second line corresponds to the calculation ignoring the stellar mass at the center.

Now, the most massive star that can exist within this core without disrupting the accretion flow must not inject thermal energy into the surrounding gas at a rate larger than $P_K$. To find the mass of this star, we can use the table of ionizing photon rates as a function of stellar mass, $\dot{N}_{\rm ion}(M_*)$, provided by \citet{Diaz-Miller98}. Specifically, $\dot{N}_{\rm ion}(M_*)$ is the number of photons with energy $e > 13.6$ eV emitted per unit time by a star of mass $M_*$. The total ionizing power emitted by the star is then $P_* = \dot{N}_{\rm ion}(M_*) \bar e$, where $\bar e$ is the characteristic energy of the ionizing photons, and can be approximated by $\bar e \approx 13.6 \eV$.

However, out of this total power, one part will be consumed in ionizing the gas, and another will be used in heating it up to the typical ionized gas temperature, with only the remaining power being available to halt the accretion flow. Denoting this fraction by  $f_{\rm h}$, and assuming that the gas is heated to a temperature $T_{\rm h}$, the radiative energy available to halt the accretion flow will be that above $13.6\eV + \kB T_{\rm h}$, which corresponds to $14.3\eV$ for $T_\text{h} = 8\times10^3 \degK$. We then estimate $f_\text{h}$ as
\begin{equation}
    f_{\rm h} = \frac{\int_{14.3\eV}^\infty B_\nu(T_{\rm eff}) \, \dif \nu}
                      {\int_{13.6\eV}^\infty B_\nu(T_{\rm eff}) \, \dif \nu},
\end{equation}
where $B_\nu(T)$ is Planck's law of blackbody radiation, and $T_{\rm eff}$ is the effective temperature of the star of mass $M_*$. Therefore, in order to compensate for the power lost to the ionization of the gas, we determine the mass of the accretion-halting star as that in Table 1 of \citet{Diaz-Miller98} whose ionizing-photon rate corresponds to a total power $P_* = P_K/f_{\rm h}$.

\begin{figure*}
\centering
\includegraphics[width=\textwidth]{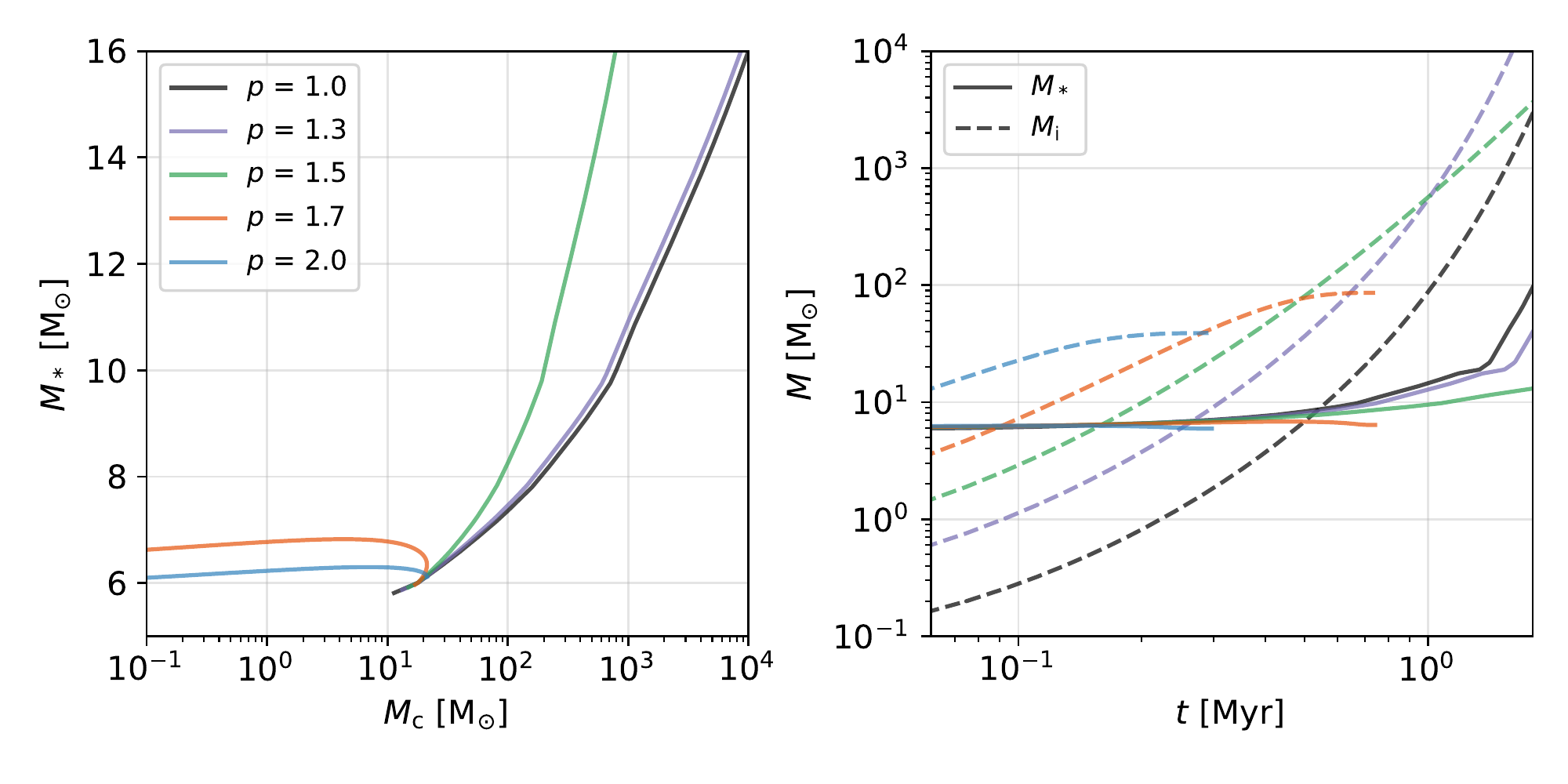}
\caption{{\it Left:} Mass of the star ($M_{*}$) required to halt, by photoionizing radiation, the accretion driven by self-gravity onto a core of mass $\Mcore$. {\it Right:} Temporal evolution of $M_{*}$ (solid lines), and of the mass that has been accumulated within an inner, characteristic ``accretion disk radius'' $r_{\rm i} = 3000 \AU$ ($M_{\rm i}$, dashed lines). 
As in Fig.\ \ref{fig:Mcore}, we have assumed an environmental density $\rho_0 = 3\times 10^3\pcc$ and an initial radius of the core $\rc(t=0)$ equal to the Jeans length at $\rho_0$. 
}
\label{fig:Mmax}
\end{figure*}

The {\it left} panel of Fig.\ \ref{fig:Mmax} shows the scaling of this accretion-destroying stellar mass, $M_*$, {\it versus} the core's mass, $M_{\rm c}$, the latter obtained by integration of eq.\ \eqref{eq:dot_Mcore_of_r}, for a range of $p$ values. 
On the other hand, the {\it right} panel of Fig.\ \ref{fig:Mmax} shows the evolution of the mass accreted onto the central region of the core with radius $r_{\rm i} = 3000 \AU$ ($M_{\rm i}(t)$; dashed lines), obtained by integrating the equation $\dot M_{\rm i} = {\cal F}(r_{\rm i})$, for the same values of $p$. Therefore, $M_{\rm i}(t)$ can be thought of as the mass of the most massive {\it possible} star (if there were no fragmentation within $r_{\rm i}$) as a function of time. In this panel, we also show the evolution of the mass of the accretion-destroying star, $M_*$ (solid lines). 

We observe that the stellar mass $M_{\rm i}$ starts smaller, but increases faster, than the accretion-destroying stellar mass $M_*$, and so eventually $M_{\rm i}$ becomes equal to $M_*$. At this time, accretion onto the core can be disrupted, ending the local star formation episode. Thus, the actual maximum possible stellar mass within a core is given by $\min[M_{\rm i}, M_*]$ at every moment in time.

\begin{figure}
\centering
\includegraphics[width=0.5\textwidth]{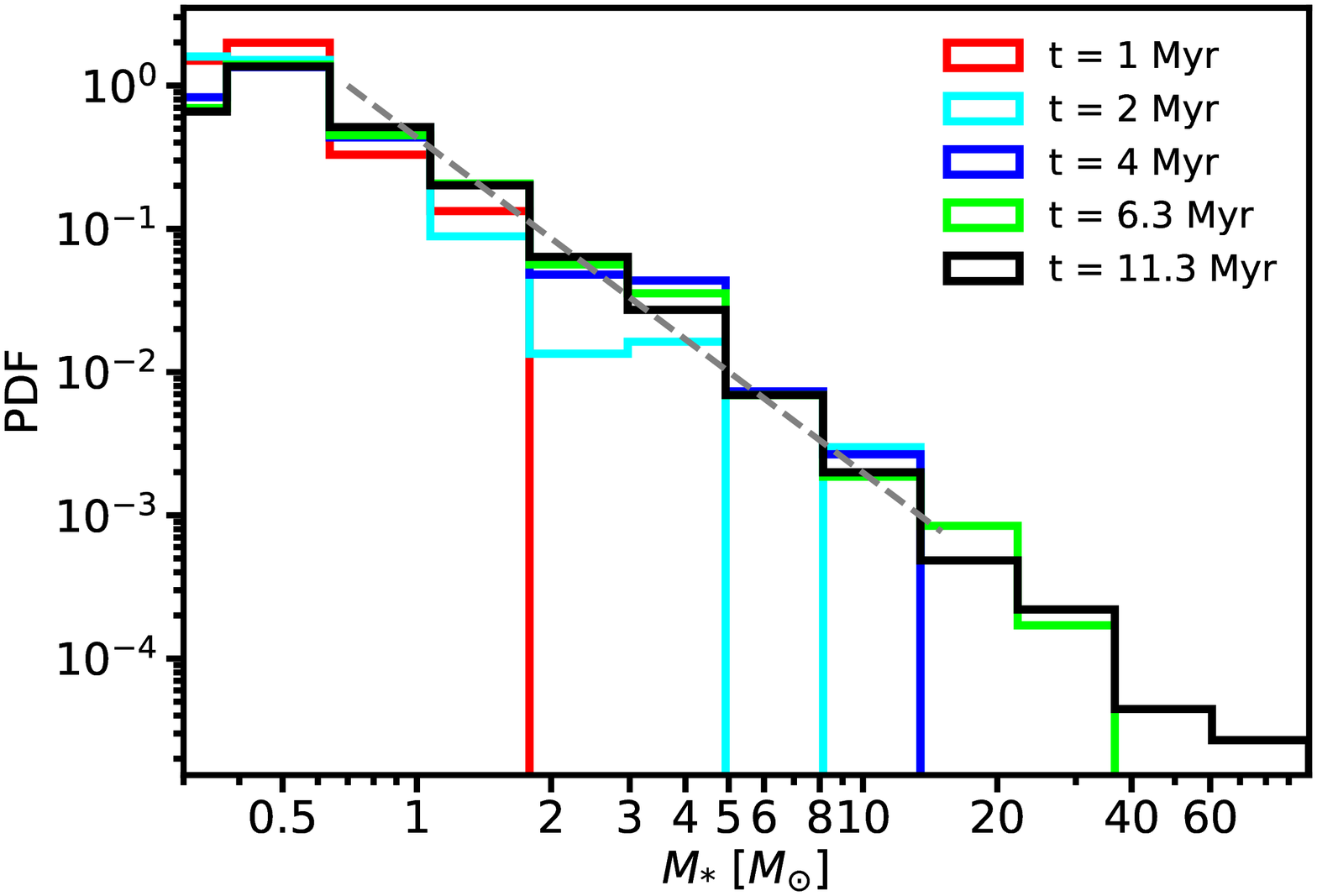}
\caption{Evolution of the mass function for one of the star-forming regions (``Group 1'') appearing in the simulation labeled ``LAF1'' of \citep{GS_VS20}, representing cluster formation in clouds undergoing global hierarchical collapse, including stellar feedback. More massive stars appear later during the evolution of the region. We show this simulation because the one without feedback (LAF0) does not produce a Salpeter slope, and produces massive stars too rapidly, although they still form later than less massive ones.}
\label{fig:massF}
\end{figure}

\begin{figure}
\centering
\includegraphics[width=0.5\textwidth]{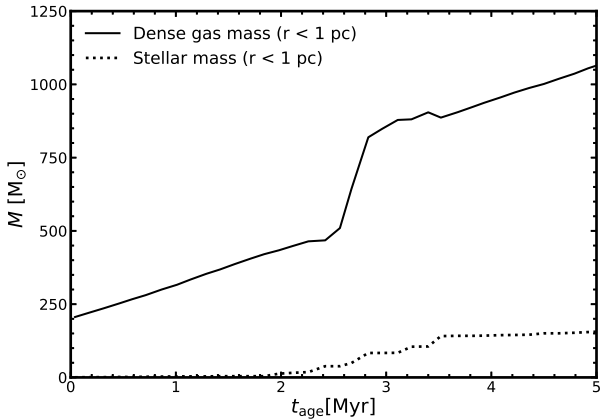}
\caption{Evolution of the stellar and dense gas mass in the inner parsec of Group 2 for the simulation without feedback (LAF0). 
}
\label{fig:evolveM}
\end{figure}

Note also that this stellar mass reaches values $\sim 10 \Msun$ within a time of the order of 1 Myr, with the precise time depending on the slope of the core's density profile. Note that this total stellar mass will be distributed among a population of collapsed objects, and therefore $M_{\rm i}$ needs to be significantly larger than $M_*$ in order to have a star of this latter mass.  Therefore, our timescales for the growth of $M_{\rm i}$ can be considered as lower limits to the true required timescale.
This can be compared to the time required in numerical simulations for massive stars to appear. For example, Fig.\ \ref{fig:massF} shows the evolution of the stellar mass distribution in the simulation LAF1\footnote{We compare to the simulation including feedback because the one without it deviates from a Salpeter slope of the stellar IMF. In any case, massive stars also appear at a later time in that simulation.} (including feedback) analyzed in \citet{VS+17} in differential (or PDF) form.\footnote{Figure 7 in \citet{VS+17} showed the same distribution in cumulative form.} It can be seen that stars of masses up to $M \lesssim 5 \Msun$ appear within 2 Myr, and a star with $M \sim 10 \Msun$ appears after $\sim 4$ Myr. At this time, the total stellar mass within 1 pc is $\sim 150 \Msun$, and the gas mass is $\sim 1000 \Msun$ (see Fig.\ \ref{fig:evolveM}). Although the model core and stellar masses shown in Fig.\ \ref{fig:Mcore} are still significantly larger than the simulation values, it is shown in App.\ \ref{app:q-profile} that a slightly more realistic case, assuming a background that is not uniform, but rather has a shallower slope than that of the core, produces numbers closer to those of the simulation.
%
Regardless, both figures \ref{fig:Mmax} and \ref{fig:massF}
illustrate the sequential appearance of more massive stars as time proceeds, implying that the region evolves from being a low-mass star-forming region to a high-mass one, as predicted by the GHC scenario \citep{VS+09}. 

It is important to note that the maximum stellar mass allowed by the feedback, $M_*$,  plotted in the two panels of Fig.\ \ref{fig:Mmax}, does {\it not} intend to predict the mass of the stars that will actually be forming within a core, since the core will undoubtedly undergo fragmentation, forming stars with a certain mass distribution, probably through the competitive accretion mechanism \citep{Bonnell+01a, Bonnell+01b}. This result should instead be interpreted as meaning that, due to the accretion flow, both the core's mass $M_{\rm c}$ and the total mass available for star formation increase with time, thus increasing the mass of the most massive possible star ($M_{\rm i}$), as well as the latter's corresponding photoionizing feedback power. When the mass of this most massive possible star surpasses the mass of the star that can disrupt the accretion flow ($M_*$), the core can stop growing, and the star-formation episode may be terminated by gas exhaustion, as inferred observationally by \citet{Ginsburg+16}.


Although our model is highly idealized and approximate, its main relevance is the implication that, when the accretion both from a clump to a core and from the core to the star(s) are taken into account, the maximum stellar mass that a core can harbor increases over time, together with the core's mass, a proof-of-concept model for the numerical result that more massive stars appear later in a cluster \citep{VS+17}, and for the observed property of clusters that the mass of the most massive star correlates with the mass of its parent cluster \citep[e.g.,] [] {Weidner_Kroupa06, Weidner+10}.

\section{Discussion} \label{sec:disc}


\subsection{Caveats} \label{sec:caveats}

Our model is certainly highly idealized, and, as a consequence, it still falls short of having a strong predictive power. Its main limitations are:

\begin{itemize}

\item {\it The restriction to spherical symmetry.} Although we have considered the possible effect of filaments of producing some flattening of the effective spherically-averaged density profile, all our accretion rates are computed assuming spherical symmetry. This assumption causes the gravitational potential to be deeper than if the same mass were to be distributed in a sheet- or filament-like geometry, causing longer infall times and smaller accretion velocities \citep{Toala+12, Pon+12}. 

\item {\it The neglect of any form of agents counteracting gravity}, such as thermal pressure or magnetic fields, or of low-mass-star feedback, that may delay the gravitational contraction.

\end{itemize}

As a consequence, our infall speeds and accretion rates should be considered as upper limits, and our evolutionary timescales as lower limits, to what may be expected in actual molecular cloud cores. Nevertheless, our model incorporates several important features observed in numerical simulations that point towards aspects of gravitational contraction that, to our knowledge, have not been previously addressed, and provides a proof-of-concept discussion of the mechanism of gravitational choking, and the importance of the density profile slope in determining accretion rates. In addition, our model compares well to order of magnitude with the numerical results of \citet{GS_VS20}.


\section{Summary and conclusions} \label{sec:concls}

In this paper we have presented a simple, idealized model, in quasi-spherical geometry, of a mechanism that can account for the observation, in numerical simulations of cloud formation and evolution from warm diffuse atomic gas, that star-forming regions grow in mass and size, and that massive stars form with a delay of a few to several megayears after the first stars begin to form. 

The model is based on two main ingredients. First, on the assumption that there is accretion both from the core to the stars and from the cloud to the core. That is, we account for the two last stages of the multi-stage accretion process predicted by the GHC scenario to be occurring in molecular clouds). Second, on the result from Paper I that the gravity-driven accretion rate in a core varies with radius in general. In a core with a power-law density profile with logarithmic slope $-p$, the accretion rate is only independent of radius when $p = 2$. For $p < 2$, the accretion rate decreases inwards, causing some of the infalling material to stagnate in the gaseous phase, increasing the core's gaseous mass. We call this process ``gravitational choking'', 
and it continues until the stellar mass at the center becomes large enough to dominate the core-to-star accretion rate, at which point the core may be depleted (Appendix \ref{app:q-profile}).

In Paper I we had furthermore shown that, under strict spherical geometry, the value $p = 2$ is an {\it attractor}, meaning that the slope evolves toward that value,  reaching it at the time when a protostar forms. However, here we have shown (Appendix \ref{sec:fils_dens_profile}) that deviations from sphericity, such as those induced by the presence of filaments feeding a hub, may allow the spherically-averaged slope to remain at values $p < 2$, thus allowing for the core's mass to grow even after stars have begun to form at the core's center, and for simultaneous growth of the core and stellar masses. 

A key element of the GHC paradigm \citep{VS+19} is the interconnection of scales through mass accretion.
Computing the accretion rate at the core's outer boundary, as well as the accretion from the core to the ``stellar region'' (defined by an inner radius within the core comparable to the estimated size of the Oort cloud), we were able to obtain the simultaneous growth by accretion of both the core's and stellar masses. The latter constitutes an upper limit to the mass of the most massive star that the core can harbor. Moreover, we also computed the mass of the star whose photoionizing radiation flux balances the power of the accretion onto the core. When the total mass in stars is large enough to produce this disrupting star, we suggest that the accretion can be halted, and the local star formation event can be terminated. In the GHC picture, the evolution of the core and the formation of massive stars with it are not defined by the core's own mass, as in the competitive accretion or the turbulent-core models, but by the mass in its environment {\it susceptible to be accreted onto the core}. So, the formation of a star capable of halting the accretion flow sets the limit of the mass available to fall onto the core.

In this way, the model implies that more massive stars require more time to form than low-mass stars, because a sufficient amount of mass must be collected at the center of the cores, and that the mass of the most massive star present in a core must correlate with the core's own mass, because the core's mass also grows while the mass available to form stars in its center increases.

Our model is, of course, highly idealized, as it assumes a spherical geometry with a single power law for the density profile, and neglects any processes opposing the collapse before the disruption of the accretion flow. The only deviation from sphericity is the consideration that filamentary structures may flatten the spherically-averaged density profile. However, our model provides a proof-of-concept that the simultaneous core-to-stars and cloud-to-core accretion processes imply a delayed formation of the more massive stars, and a correlation between the mass of the most massive star and that of its parent core. The latter suggests that a correlation between the mass of the most massive star and that of its host cluster should exist as well.

Note, however, that the delayed formation of the massive stars does not imply that the less massive stars are all older than the more massive ones. This is because the star formation rate also increases, and so most of the low-mass stars are coeval with the more massive ones \citep{VS+17}. However, a small population of old, low-mass stars is expected to exist in the cluster in addition to the majority of young, low- and high-mass stars. The main implication of our model is that the distribution of stellar masses in a star-forming region extends to ever larger masses as time proceeds, until the local episode of star formation is halted by the stellar feedback.

Finally, our model can be considered as a time-dependent alternative to the {\it turbulent core} model of \citet{McKee_Tan03}, which was based on the assumption of turbulent support of a massive core, and has triggered intense searches for prestellar massive cores. Instead, our model generally predicts that, by the time massive stars begin to form in a clump or core, a significant number of low-mass must have already formed.







\section*{Acknowledgements}

We thankfully acknowledge a very useful and constructive report by an anonymous referee, and insightful comments by William Henney and Sally Oey. EVS acknowledges financial support from UNAM-PAPIIT grant IG100223. GCG acknowledges support from UNAM-PAPIIT grant IN10382. 

\section*{Data availability}
The data underlying this article will be shared on reasonable request to the corresponding author.



\bibliographystyle{mnras}
\bibliography{refs} 



\appendix

\section{Simulations} \label{sec:methods}

In \citet{GS_VS20}, two simulations of converging flows in the warm Galactic atomic medium were considered, one without stellar feedback, labeled LAF0, and one with a sub-grid prescription for emulating the UV ionizing feedback from massive stars, extended down to masses $\sim 1 \Msun$, labeled LAF1. The numerical box was 256 pc on a side, at a maximum resolution of 0.0625 pc. These simulations, first presented in \citet{Colin+13}, used a stochastic star formation prescription that allowed the sink particles to have stellar masses and with a Salpeter \citep{Salpeter55} slope in the case with feedback, making them the first simulations at the giant molecular cloud scale with stellar-mass sink particles. We refer the reader to that paper for details on the simulations.

In Figs.\ \ref{fig:profile_ART}, \ref{fig:massF}, and \ref{fig:evolveM} this paper, we refer to ``clumps'', which are defined as spherical regions centered at the center of mass of the stellar clusters forming in the simulations, and of the radii indicated in the figures.

\section{The effect of filaments on a core's density profile} \label{sec:fils_dens_profile}


The results from section \ref{sec:retained_mass_frac} apply to a spherical collapse flow. However, in reality, star-forming regions in general are far from spherical, and at present it is agreed that they consist of ``filament-hub'' systems, in which the central, approximately spherical hubs are ``fed'' with fresh gas by a network of filaments \citep [e.g.,] [] {Myers09, Schneider+10, Kirk+13, FernandezL+14, JimenezS+14, Peretto+14, Ginsburg+16, Wyrowski+16, Juarez+17, Gong+18, ChenV+19}. The filaments are known to have a nearly uniform transverse column density (and presumably linear mass density) along their length \citep[e.g.,] [] {Andre+14}, and so one can expect that the spherically-averaged profile will be shallower than if the filaments were not present.

\begin{figure*}%
\centering
\includegraphics[width=0.9\textwidth]{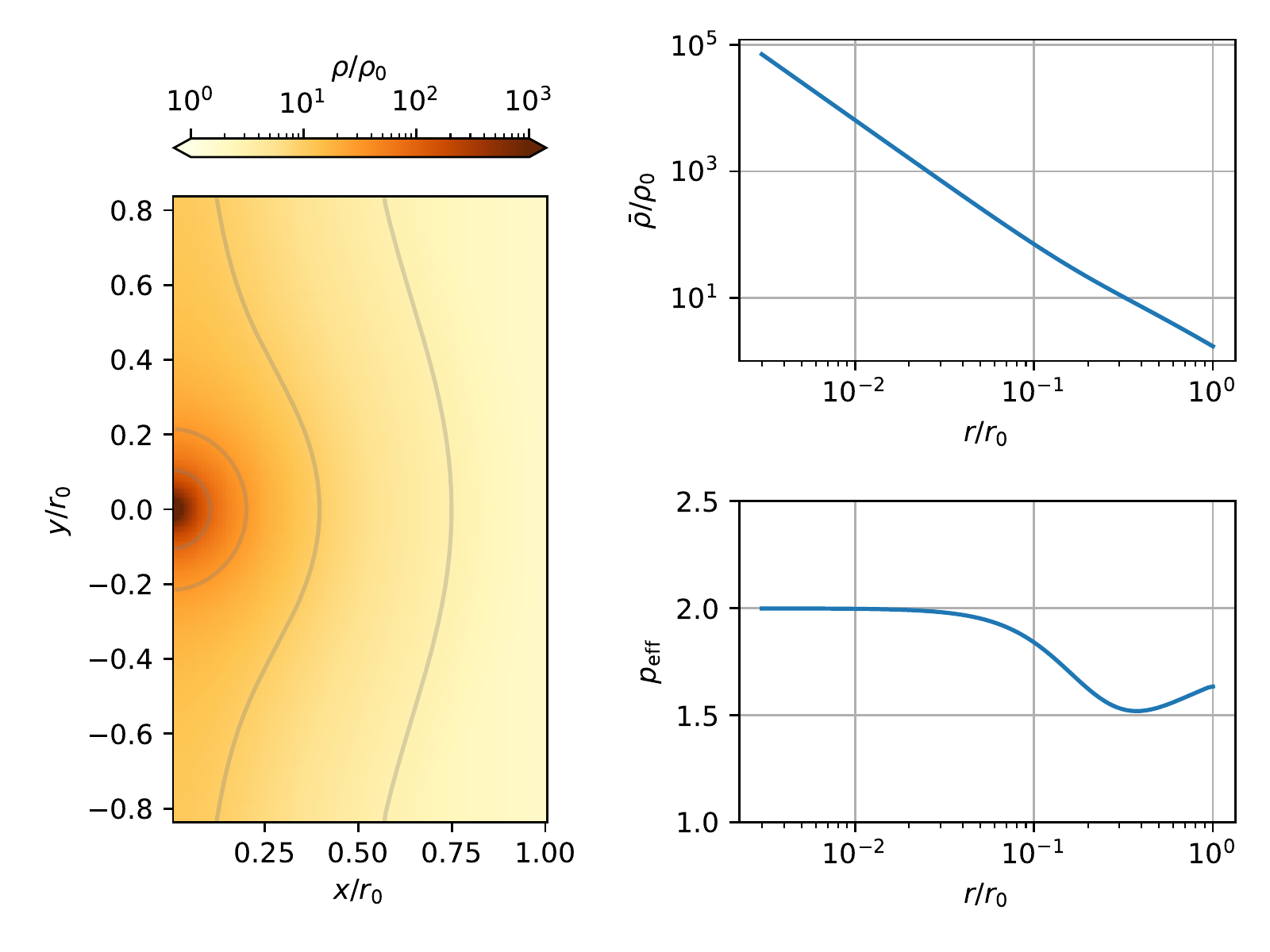}
\caption{Spherically averaged density profile of a hub-filament system. The core's density profile is given by eq. \eqref{eq:dens_prof}, while the embedded filament has a density $\rho(R) = \rho_{\rm f} / [1+(R/R_0)^2]^{p/2}$. In this example, $\rho_{\rm f}/\rho_0 = 10$ and $R_0/r_0=0.3$, while $p=2$ for both distributions.
{\it Left:} Density distribution for the core+filament system.
{\it Right top:} Spherically averaged density profile, $\bar\rho$, of the superposed distributions. {\it Right bottom:} Effective profile slope,
$p_{\text{eff}} = - \dif\log(\bar{\rho})/\dif\log(r)$. As $r$ increases, the filament contribution keeps the density above the value corresponding to the clump distribution, thus flattening the profile.}
\label{fig:mean_density}
\end{figure*}

Consider the case of a core embedded in a filament with a Plummer-like radial density profile. 
Thus, the total density of the distribution, $\rho_{\rm T}$, is
\begin{equation}
    \rho_{\rm T} = \rho_0 \left(\frac{r}{r_0}\right)^{-p} +  \frac{\rho_{\rm f}}{[1+(R/R_0)^2]^{p/2}},
\end{equation}
with $R = r \sin\theta$ being the cylindrical radial coordinate, $\theta$ the angle with respect to the filament's axis, $\rho_{\rm f}$ the filament axial density, and $R_0$ a radial scale. The configuration is illustrated in the left panel of Fig.\ \ref{fig:mean_density}, and its spherically averaged density profile can be computed numerically. The mean density profile, $\bar\rho(r)$, and its effective logarithmic slope are also shown in the top- and bottom-right panels. When $r < r_0$, the core density dominates and $\bar\rho$ reflects the $r^{-p}$ profile. But, as the filament density becomes important, the profile flattens significantly.

It should be noted that this flattening of the slope is important at the {\it dynamical} level. Since the accretion flow is driven by the gravitational potential of the mass distribution of the hub-filament system, the fact that the spherically-averaged mass distribution has an effective slope $p<2$, implies that we can expect the gravitational choking mechanism to continue operating, and the core can continue to increase its mass, regardless of whether protostar formation has already begun in the system, as long as a filamentary component remains in the system.


\section{Timescales for profile steepening and radius growth} \label{sec:app_acc_times}

\begin{figure}[h]%
\centering
\includegraphics[width=0.5\textwidth]{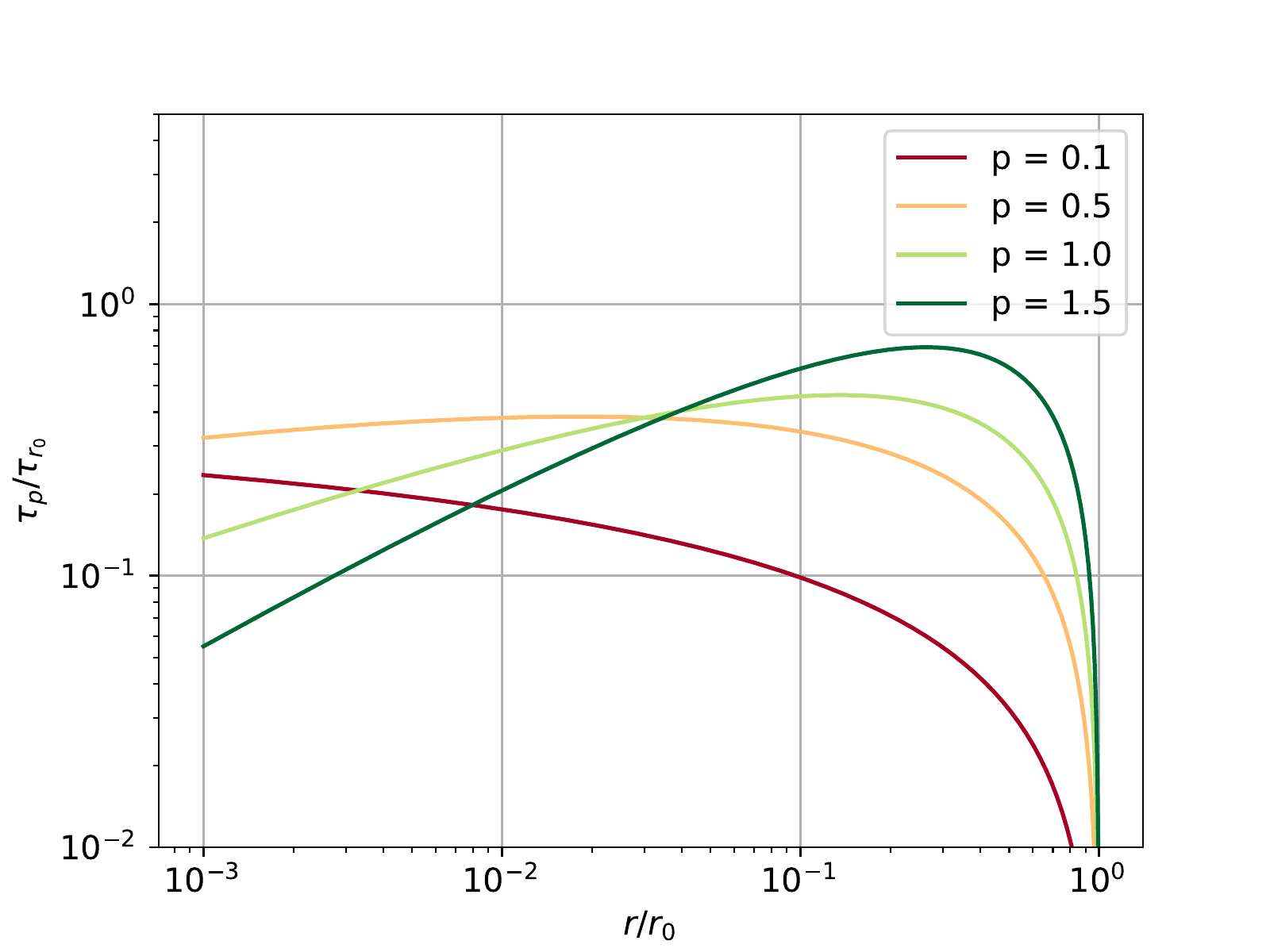}
\caption{Ratio of the timescale for growth of the density slope $p$ ($\tau_p$) to that for core-size increase ($\tau_{r_{\rm c}}$).
The timescale for increasing $p$ is generally shorter than that for increasing $r_0$, and so the density distribution steepens faster that the core expands.}
\label{fig:acc_times}
\end{figure}

The core's mass may grow through steepening of its density profile (driven by accretion onto the core) and/or by increasing its size.
The timescale for the former process is $\tau_p = p / (\dif p/\dif t)$, with $\dif p/\dif t$ given by eq. \eqref{eq:dpdt}, while the latter is found by setting the accretion rate, eq. \eqref{eq:accr_rate}, evaluated at the core boundary $r_0$, equal to the time derivative of the core's mass, eq. \eqref{eq:Mdot_core}, and defining the timescale for core expansion as $\tau_{r_0} = r_0 / (\dif r_0/\dif t)$.
The ratio of these timescales is given by
\begin{equation}
    \frac{\tau_p}{\tau_{r_0}} = \frac{\sqrt{2}p}{3} \left( \frac{1-p/3}{1-p/2} \right)
      \left[ \frac{-\ln(r/r_0)}{(r/r_0)^{-p/2}} \right].
\end{equation}
Figure \ref{fig:acc_times} shows this ratio as a function of radius.
It is seen that, throughout most of the core's volume, the density distribution first steepens and then the core expands. In addition, recalling that the slope becomes stationary at $p=2$, and that at that time a collapsed object forms, the result $\tau_p < \tau_{r_0}$ suggests that $p$ increases during the prestellar stage, while $r_0$ increases during the protostellar stage. 

\section{A non-uniform medium around the core}
\label{app:q-profile}

\begin{figure}
    \centerline{
      \includegraphics[width=0.5\textwidth]{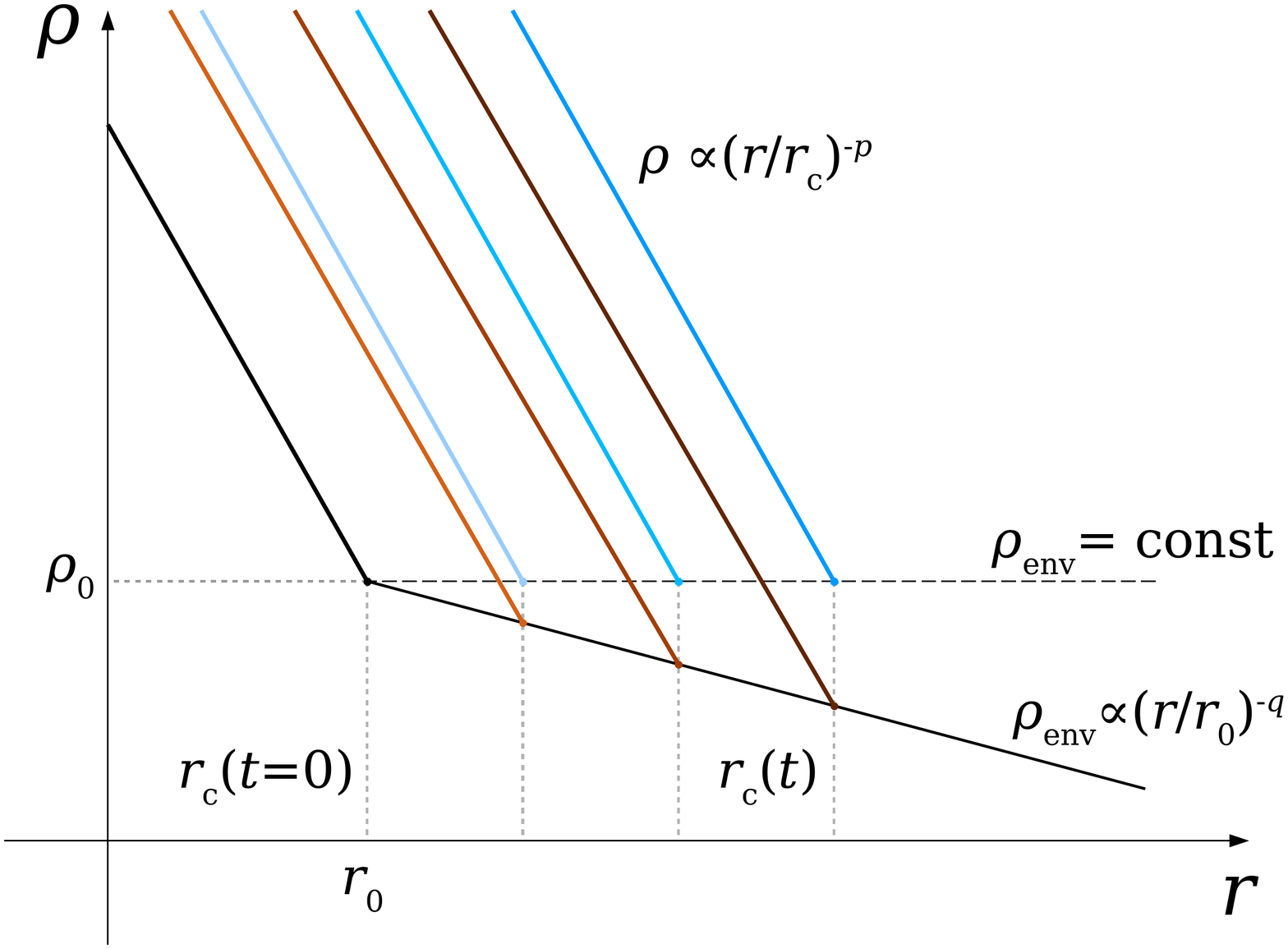}
    }
    \caption{Schema for the density profile of the core surrounded by a decreasing-density environment. When growing into a density distribution $\rho \propto r^{-q}$, the core's profile ({\it brown lines}) is lower than a similar size core growing in a constant density environment ({\it blue lines}).}
    \label{fig:perfil_q}
\end{figure}

The cores obtained from the model discussed in sec. \ref{sec:growth} develop masses that grow unrealistically. One of the reasons for such a growth is the assumption that the core is surrounded by an infinite environment of uniform density. We may relax this assumption by assuming that the core's environment also follows a power-law density profile, 
\begin{equation}
\rho_{\rm env}(r) = \rho_0 \left(\frac{r}{r_0}\right)^{-q}.
\label{eq:perfil_q}
\end{equation}
This is actually a reasonable assumption, as the typical density of molecular cloud cores at the $0.1$-pc scale is $\sim 10^4$ cm$^{-3}$, while the mean density of molecular clouds at the $10$-pc scale is $\sim 100$ cm$^{-3}$, in fact suggesting $q \sim 1$.

\begin{figure*}
\centering
\includegraphics[width=1\textwidth]{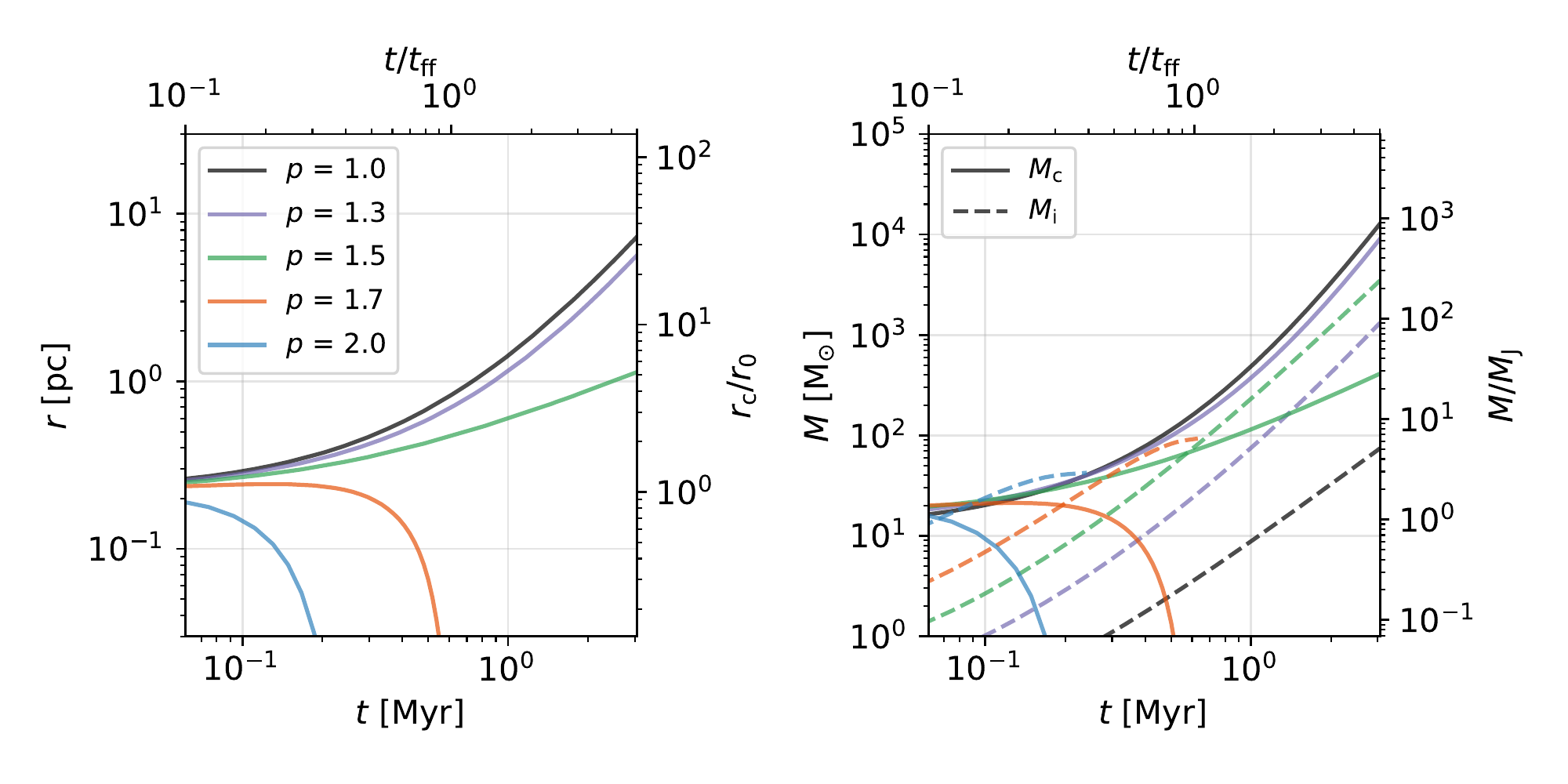} \\
\caption{Core size ($\rc$, {\it left}) and core and stellar mass ($M_{\rm c}$ and $M_{\rm i}$, {\it right}) evolution for the case of the core growing into a decreasing density environment with $q=1$. This figure is equivalent to fig. \ref{fig:Mcore} and uses the same value for $\rho_0$ and $\ri$.}
\label{fig:Mcore_q1}
\end{figure*}

In order to maintain a continuous density distribution, the core's density profile must be modified from eq. \eqref{eq:dens_prof} to
\begin{equation}
\rho(r) = \rho_c \left(\frac{r}{\rc}\right)^{-p},
\label{eq:dens_prof_modified}
\end{equation}
with $\rho_{\rm c} = \rho_{\rm env}(\rc)$ (see fig. \ref{fig:perfil_q}). Thus, as the core's radius grows, the density profile of the core remains lower than that of a similar-sized core growing in an uniform density environment, as those modeled in Sec.\ \ref{sec:growth}. Also, the growing core accretes material of ever-diminishing density. The mass flux at the core's boundary will be given by,
\begin{equation}
    \Fcal(\rc) = 4 \uppi \rc^2 \rho_{\rm env}(\rc) v_{\rm inf},
    \label{eq:Flux_rc_mod}
\end{equation}
with $v_{\rm inf}$ still given by eq. \eqref{eq:vgrav_tot}. This new $\mathcal{F}(\rc)$, and $\mathcal{F}(\ri)$ from sec. \ref{sec:growth}, are integrated in time to obtain the core and stellar mass evolution in time (see Fig.\ \ref{fig:Mcore_q1}, which is equivalent to Fig.\ \ref{fig:Mcore}, but or a non-uniform environment). Although the core's gas and stellar mass are still large, they are a few orders of magnitude smaller than that of the core growing in a uniform density, and compare better to those of the numerical simulation (Fig.\ \ref{fig:evolveM}). On the other hand, the main features discussed in sec. \ref{sec:growth} are still present, like the fact that the critical slope of density profile required for the core to be emptied onto the stellar region remains at $p = 3/2$.

\bsp	
\label{lastpage}
\end{document}